\documentclass[numberedappendix]{emulateapj}
\usepackage{amssymb}

\def\deg {{\hbox{$^\circ$}}}
\def\eq#1{\begin{equation} #1 \end{equation}}
\let\mic=\micron

\slugcomment{Submitted 2005 June 07; Accepted in ApJ 2005 September 08}

\shorttitle{Optically thin halos around inner disk regions}
\shortauthors{Vinkovi\'{c} et al.}

\begin{document}

\title{Near infrared and the inner regions of protoplanetary disks}

\author{Dejan Vinkovi\'{c}}
\affil{Institute for Advanced Study, School of Natural Sciences,
       Einstein Drive, Princeton, NJ 08540; dejan@ias.edu}

\author{\v{Z}eljko Ivezi\'{c}}
\affil{Department of Astronomy, University of Washington,
       Seattle, WA 98195; ivezic@astro.washington.edu }

\author{Tomislav Jurki\'{c}}
\affil{Department of Physics, Faculty of Science, University of Zagreb,
       Bijeni\v{c}ka 32, HR-10002 Zagreb, Croatia}

\and

\author{Moshe Elitzur}
\affil{Department of Physics and Astronomy, University of Kentucky,
       Lexington, KY 40506; moshe@pa.uky.edu}

\begin{abstract}
We examine the ``puffed-up inner disk'' model (Dullemond, Dominik
\& Natta 2001), proposed for explaining the near-IR excess
radiation from Herbig Ae/Be stars. Detailed model computations
show that the observed near-IR excess requires more hot dust than is
contained in the puffed-up disk rim. The rim can produce the
observed near-IR excess only if its dust has perfectly gray
opacity, but such dust is in conflict with the observed 10$\mu$m
spectral feature. We find that a compact ($\sim$ 10 AU) tenuous
($\tau_V \la 0.4$) dusty halo around the disk inner regions
contains enough dust to readily explain the observations.
Furthermore, this model also resolves the puzzling relationship
noted by Monnier and Millan-Gabet (2002) between luminosity and
the interferometric inner radii of disks.
\end{abstract}


\section{Introduction}

Processes in the immediate vicinity of young pre-main-sequence stars influence
the initial stellar evolution and the formation of terrestrial planets. Since
small scales of several AU around a star are difficult to resolve, we still
lack a clear understanding of processes such as disk accretion, launching of
bipolar jets and winds, and dynamics and reprocessing of dust in the inner hot
disk regions. The dust geometry is one of the basic ingredients needed for
constraining theoretical models of these processes. Traditionally, this
geometry has been deduced from the spectral energy distribution (SED), which is
dominated at infrared wavelengths by dust emission.

A widely popular geometrical description is the two-layered flared disk model
developed by \citet{CG97} (CG hereafter). The model gives a simple method for
estimating the flux from the optically-thin surface layer of an optically-thick
disk directly exposed to the stellar radiation, and from the colder disk
interior heated by the warmer surface. The simplicity of the method, together
with evidence for the existence of disks based on radio imaging, made this
model a dominant description of T Tau and Herbig Ae/Be stars (intermediate-mass,
1.5$\la M_*/M_\odot\la$10, counterparts of T Tau; Haebes hereafter).

Although the CG model successfully explains the observed SEDs,
advances in imaging techniques revealed shortcomings of this
model. Analyzing images at scattering and dust emission wavelengths,
\citet{MIVE} concluded that disks alone cannot explain the imaging
observations, at least not for some Haebes.  Instead they modeled the
SED with an optically thin halo surrounding an optically thick disk,
and emphasized that only multi-wavelength imaging can distinguish
between this and the CG model. Subsequent detailed modeling of imaging
data in numerous systems revealed the existence of dusty halos around
the putative flared disks \citep{HL_Tau,GM_Aur,HV_TauC}. The
inadequacy of the SED as the sole analysis tool in determining the
geometry was further demonstrated by \citet{Vinkovic03} (V03
hereafter). They showed that the mathematical expression for the SED
calculation in the CG model can be transformed into that for the
halo-embedded-disk and vice versa. This has far reaching consequences
for all studies based solely on SEDs. If not supported by imaging at
various wavelengths, SED models can lead to erroneous conclusions
about the spatial distribution of dust.

The disk inner region in Haebes (within $\sim$10 AU from the star) proved
to be more complicated than the original CG model. Thermal dust emission from
this region peaks at short wavelengths, creating a near IR bump ($1\mu
m<\lambda<8\mu m$) in the SED of many Haebes \citep{Hillenbrand}.
\citet{Chiang01} noticed that the CG model did not produce enough near IR flux
to explain the bump. This implies that the disk flaring, which increases the
emitting volume of the optically thin disk surface, is too small at the inner
radii. Since the disk geometry is constrained by vertical hydrostatic
equilibrium, an additional hot dust component is required for explaining the
near IR bump. To solve this problem, \citet{DDN} (DDN hereafter) proposed to
modify the CG geometry without introducing an additional component. They noted
that the disk vertical height is increased (puffed up) at its inner rim because
there the disk interior is directly exposed to the stellar radiation and hotter
than in the CG model at the same radius. The rim is the hottest region of the
disk and with its increased size it is possible to boost the near IR flux. This
puffing of the rim is equivalent to the disk extra flaring that was identified
by \citet{Chiang01} as missing in the CG model.

\begin{figure*}
 \epsscale{.7} \plotone{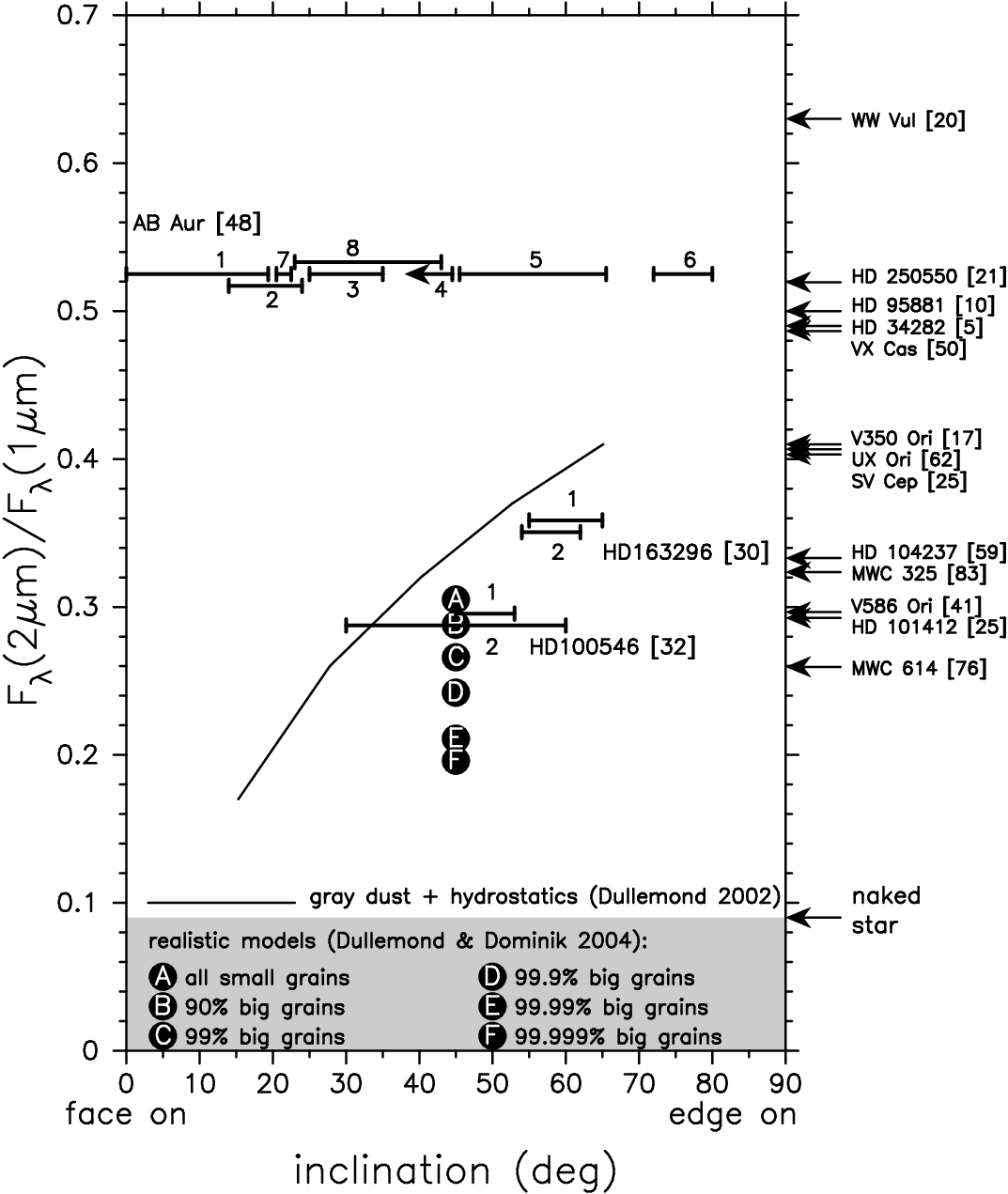}
 \caption{ \label{F2F1_Dullemond}
Strength of near IR bump in Herbig Ae stars. The ``naked star" arrow
on the right axis marks the strength for a 10,000 K black-body
spectrum. Other arrows indicate the data (see \S\ref{section2}
for references) for systems with unknown disk inclination
angles. These angles were estimated for three stars, as indicated: HD
163296 --- 1: \citet{Grady00}, 2: \citet{MS97}; HD 100546 --- 1:
\citet{Grady01,Augereau}, 2: \citet{Liu03}; AB Aur --- 1:
\citet{Eisner04} and \citet{Eisner03}, 2: \citet{Semenov04}, 3:
\citet{Fukagawa_AB_Aur}, 4: \citet{Grady99}, 5: \citet{Liu04}, 6:
\citet{MS97}, 7: \citet{Corder}, 8: \citet{Pietu}.  Stellar
luminosities in units of $L_{\odot}$ are indicated together with the
stellar name.  The {\it solid line} is a 2D radiative transfer model
of puffed up gray dust wall by \citet{Dullemond2002}.  The {\it filled
circles with letters} (realistic models, inclination $i$ = 45\deg\ )
are 2D radiative transfer models by \citet{DD04} that include a
mixture of big and small dust grains.  The rim height in all these
models is calculated directly from the equation of vertical
hydrostatics.  While the gray dust model can explain the majority of
data, the realistic models of puffed up rim fail to explain the data
even when 99.999\% of dust mass is in big (gray) grains.
}
\end{figure*}

Evidence in support of the DDN model was garnered from SED modeling of
a large sample of Haebes \citep{Dominik}, and recently also of T Tau
stars \citep{Muzerolle}.  Still, the inner disk geometry remains
controversial.  Recent advances in near IR interferometry provide
imaging data of this region, and the first results from a large sample
of Haebes show that many of these objects appear close to circular
symmetry \citep{Millan-Gabet-survey, Eisner04}. This is an unusual
result if disk inclinations are random.  It also creates a new set of
problems when interpreted as almost face-on disks because that often
conflicts with outer disk inclinations derived from other imaging
observations (HST, radio). This is difficult to accommodate in
disk-only models, but is easily explained by halo-embedded disks
(V03).

In this paper we reexamine the DDN model and the theoretical approach
behind it, and identify some unresolved issues in its description
of the rim emission. We employ exact radiative transfer calculations
of the rim's brightness and show that the concept of puffed up rim
requires some fine tuning of the model parameters in order to produce
enough flux to explain the observations (e.g. the dust must be
perfectly gray).

Various independent observations indicate the existence of compact
halos ($\sim$10 AU) around the disk inner regions (see V03 and
references within), and we find that such halos readily explain also
the observed near-IR excess. Furthermore, the halos also resolve the
puzzling relationship noted by \citet{Monnier} between luminosity and
the interferometric inner radii of disks.

\section{Emission from the inner wall}
\label{section2}

A distinct feature of the near IR bump is its anomalously high
flux $F_{\lambda}$ as compared with the stellar emission. To
quantify this effect we introduce the flux ratio
$F_{2\mic}/F_{1\mic}$ as a measure of the strength of the near IR
bump; this ratio increases when the prominence of the bump becomes
larger. These wavelengths are chosen because the 2$\mu$m flux is
dominated by the dust, while the 1$\mu$m flux is dominated by the
star.

Figure \ref{F2F1_Dullemond} summarizes the observed values of the
$F_{2\mic}/F_{1\mic}$ flux ratio for a sample of well-observed stars,
with data compiled from the following references:
\citet{data-low70,data-gillett71,data-strom72,data-allen73,data-cohen73a,
data-cohen73b,data-cohen73c,data-glass74,data-cohen75,
data-cohen76,data-kolotilov77,data-cohen80,data-bouchet82,data-lorenzetti83,
data-tjin84,data-kilkenny85,data-the85,data-olnon86,data-berrilli87,
data-strom89,data-hu89, data-lawrence90,data-fouque92,data-berrilli92,
data-hutchinson94,data-li94,data-prusti94,data-sylvester96,data-garcia97,
data-malfait98,data-herbst99,data-ancker00,data-winter01} and
A. S. Miroshnichenko (2005, private
communication).

Dust extinction at 1$\mu$m is larger than at 2$\mu$m and could enhance the
observed strength of the near IR bump by $\sim$20\% for $A_V = 1$, therefore
only objects with $A_V \la 1$ were considered. Since the reddening
correction is negligible, the uncorrected data displayed in the figure
represent the true range of near IR bump strength in Herbig Ae stars. The
underlying stars of all objects have temperatures of about 10,000 K, which
gives $F_{2\mic}/F_{1\mic}$ = 0.09. Yet in all objects this ratio exceeds 0.25,
reflecting a large NIR excess from hot dust emission (Hillenbrand et al. 1992).
The luminosity of each object is displayed together with its name in figure
\ref{F2F1_Dullemond} and it ranges from $\sim$5$L_{\odot}$ to
$\sim$80$L_{\odot}$. Luminosity does not show any correlation with the near IR
bump strength, reaffirming our conclusion that these data can be used as a
general description of the near IR bump strength in Herbig Ae stars.

\subsection{General Description of the Rim Emission}
\label{general}

\begin{figure}
  \plotone{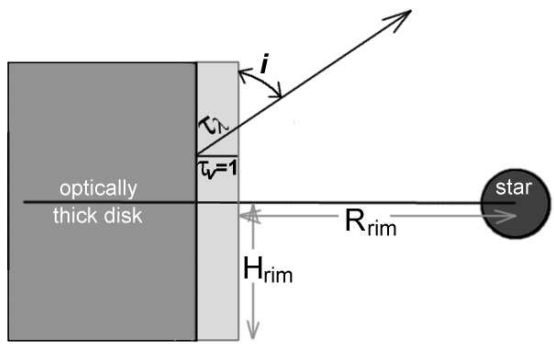}
 \caption{\label{DDN_sketch}
Sketch of the puffed up inner disk wall (see \S\ref{general}
and \S\ref{approximate} for details).
}
\end{figure}

At the inner rim, gas that is typically part of the disk cold interior
becomes directly exposed to the stellar radiation and expands to higher
scale heights.  According to DDN, emission from such a puffed-up rim
can explain the near IR bump in the spectrum of Herbig Ae/Be
stars. The rim geometry is sketched in figure \ref{DDN_sketch}. The
rim is modeled as a cylinder of radius $R_{rim}$ and height 2$H_{rim}$
centered on the star. The basic assumption of this model is that the
rim is optically thick in the near IR and shorter wavelengths. This
maximizes the rim energy output.

The original DDN model \citep{DDN} successfully explained the data,
but it was based on an approximate treatment of the rim height and
emission.  More realistic models were calculated by the authors of the
DDN model in their subsequent work.  \citet{Dullemond2002} used a 2D
radiative transfer model for gray dust combined with the hydrostatic
equilibrium.  The obtained near IR bump strength is shown in figure
\ref{F2F1_Dullemond} (solid line). The maximum strength is still too
low to explain all the data, but it can accommodate the majority of
observed near IR bump strengths.

A dramatic reduction of the DDN model efficacy happens when a mixture
of small and big grains is introduced. \citet{DD04}(DD04 hereafter)
combined 2mm (big, gray grains) and 0.1$\mu$m (small) grains in
various ratios and performed 2D radiative transfer calculations
coupled to the equation of vertical hydrostatics and dust
settling. The model fails to explain the data even when 99.999\% of
the dust mass is in big grains (see figure \ref{F2F1_Dullemond}).  The
behavior of this result is unexpected; decrease in the small grain
population leaves more gray dust grains in the mix, which should move
the whole solution closer to the gray dust result of
\citet{Dullemond2002}.

A closer inspection of obtained results shows that the temperature of
big grains in such a multi-grain mixture is lower than in the pure
gray model. In a mixture, both small and big grains absorb a fraction
of the local energy density and participate in providing the local
diffuse heating.  But, as shown in the next section (see also equation
\ref{psi} in the appendix), small grains are a very inefficient source
of local diffuse heating, resulting in less efficient heating of the
big grains then in the pure gray model. With such a temperature
decrease, the vertical hydrostatic equilibrium cannot produce disk
puffing comparable to the gray model. While in the gray model the
puffed-up disk rim height is close to $H_{rim}/R_{rim}=0.2$,
multi-grain models have only $H_{rim}/R_{rim}< 0.15$. Since the
observed rim emission scales with rim height, this is the major reason
behind the failure of the multi-grain models to explain the data. The
presented model with the lowest fraction of small grains yields the
largest discrepancy because it suffers the largest reduction in the
small grain contribution to gas heating and rim puffing while still
having enough small grains to suppress heating of the big grains. In
quantifying this effect, the mass ratio between big and small grains
that DD04 used is not the most illustrative choice. A more appropriate
quantity would be the ``equivalent'' grain size of the grain
mixture. In the case of two grain populations with sizes $a_{big}$ and
$a_{small}$ and fractional number densities $X_{big}$ and $X_{small}$
(such that $X_{big}+X_{small}=1$), the average grain size obeys
 \eq{
  \langle a^2\rangle = X_{big}a_{big}^2 + X_{small}a_{small}^2\,.
 }
The number fractions can be deduced from the reported parameters of
the DD04 models: fixed inner and outer disk radius, fixed total disk
mass, and total dust mass in big and small grains. The model with
99.999\% of the dust mass in big grains has only $X_{big}=1.25\times
10^{-8}$, and $X_{small}\sim 1$, yielding $\langle a^2\rangle^{1/2}\sim
0.25\mu m$. This grain size is too small to be considered equivalent
to the gray dust model.

It is important to note that this equivalent grain is just an indicator of the
overall solution and {\it cannot} be used as a general replacement (average or
synthetic) grain for the radiative transfer calculation.  As already shown by
\citet{Wolf}, the approximation of an averaged single grain as a replacement
for a dust mixture breaks down at the surface of a dust cloud (or in this case
the rim surface).  A more detailed study of multi-grain disk models will be
presented in a separate publication, while in the next sections we explore the
limits of possible DDN model applicability in the context of single dust
grains.

\begin{figure*}
 \epsscale{.7} \plotone{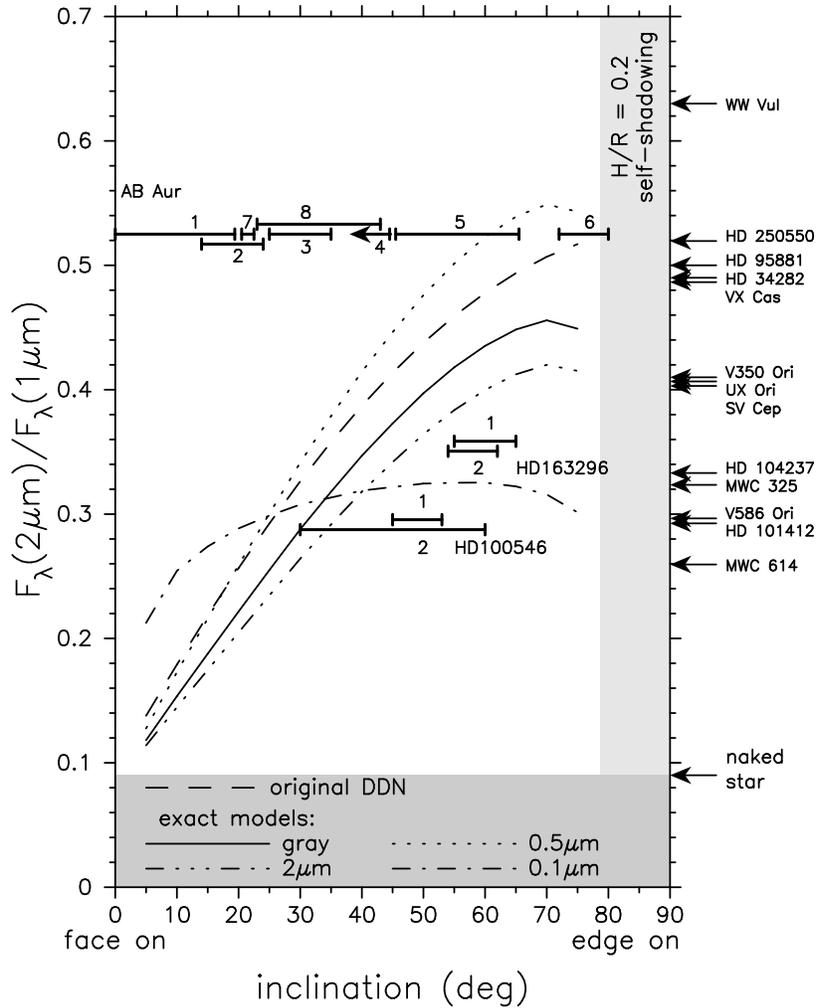}
 \caption{
Same as figure \ref{F2F1_Dullemond}, but for single-size grain models with
$H_{rim} = 0.2R_{rim}$. The shaded areas are regions without flux contribution
from the rim, because of either the absence of dust or rim self-shadowing. The
{\it dashed line} is the original DDN model \citep{DDN}; all other lines show
the results of exact 2D radiative transfer calculations (see
\S\ref{exact_DDN}).
}\label{LELUYAmodels}
\end{figure*}

\subsection{Approximate Solution for the Rim Emission}
\label{approximate}

Denote by $R_*$ and $T_*$ the stellar radius and temperature, respectively. At
distance $d$ and direction $i$ where the star is free of rim obscuration (see
figure \ref{DDN_sketch}), the overall observed flux at wavelength $\lambda$ is
$(R_*/d)^2\pi B_\lambda(T_*) + F^{rim}_{\lambda}(i)$. If $I^{rim}_{\lambda}(i)$
is the rim surface brightness in the observer's direction then
\eq{\label{Fcorrect}
 F^{rim}_{\lambda}(i) = {1\over d^2} I^{rim}_{\lambda}(i)\,4H_{rim}R_{rim}\sin i.
}
Here the cylindrical visible surface is replaced with a flat rectangle. This
approximation maximizes the flux since curvature decreases the projected area
of portions of the visible surface, reducing the observed flux. We also assume
that the stellar illumination is perpendicular to all portions of the rim. This
too maximizes the observed flux.

The observed rim flux in equation \ref{Fcorrect} is determined by the rim
height, surface brightness and radius. Our 2D radiative transfer calculations
described in \S\ref{exact_DDN} confirm that the rim emission is indeed
proportional to the rim height, therefore we maximize the rim emission in this
study by using $H_{rim} = 0.2R_{rim}$, the maximum height allowed before the
rim starts to shadow large portions of the disk (DDN). The solution for any
other rim height can be derived from our models by a simple scaling of the rim
emission.

The surface brightness of a gray dust rim can be approximated with
$B_{\lambda}(T_{rim})$, where $T_{rim}$ is the dust sublimation temperature.
The description of a non-gray surface must take into account the spectral
variation of optical depth of the emitting optically-thin surface layer. This
was done by \citet{CG97}. According to their model, the surface layer vertical
optical thickness is unity at visual (a characteristic wavelength of the
stellar radiation absorption) $\tau_V = 1$, therefore at all other wavelengths
it is $\tau_{\lambda} = \sigma^{abs}_{\lambda}/\sigma^{abs}_{V} \equiv
q_{\lambda}$. The rim emits at near IR where $q_{\lambda} < 1$ (the dust NIR
opacity is smaller than at visual), thus the surface layer is optically thin at
these wavelengths and its emission is reduced accordingly. Therefore, the rim
surface brightness becomes $\simeq q_{\lambda}B_{\lambda}(T_{rim})/\sin i$ and
the observed rim flux is
\eq{\label{Fnongray}
  F_{\lambda} \simeq {4\over d^2} B_{\lambda}(T_{rim})\,
                            {H_{rim}\over R_{rim}}R^2_{rim} \times
    \cases{q_{\lambda}  &   non-gray dust \cr
                                                           \cr
           \sin i       &   gray dust \cr
    }
}
This result shows that {\it a non-gray rim creates a smaller IR excess than a
gray opacity rim}. In addition, {\it non-gray opacity removes the angle
dependence from the rim emission} (we expect this approximation to break down
at very small inclination angles where $q_{\lambda}\sim  \sin i$).

The rim radius is derived from radiative equilibrium, which gives
\citep[e.g.][]{IE97}
 \eq{\label{Rcorrect}
 R_{rim} = \frac12 R_* \left(T_*\over T_{rim}\right)^{2}
 \left[ {\bar{\sigma}(T_*)\over \bar{\sigma}(T_{rim})} \psi
       \left(1+{H_{rim}\over R_{rim}}\right) \right]^{1/2}
 }
Here $\bar{\sigma}(T)$ is the Planck average of $\sigma^{abs}_{\lambda}$ at
temperature $T$, $\psi$ describes the correction for diffuse heating from the
rim interior\footnote{Note that \cite{IE97} used $\Psi =
{\bar{\sigma}(T_*)\over \bar{\sigma}(T_{rim})} \psi$ } and $1+H_{rim}/R_{rim}$
is a correction (described by DDN) for self-irradiation from the other side of
the rim. In appendix \ref{appendixA} we derive an approximate solution which
shows that gray dust, with ${\bar{\sigma}(T_*)/\bar{\sigma}(T_{rim})} = 1$, has
$\psi\sim 4$ and that non-gray dust, with
${\bar{\sigma}(T_*)/\bar{\sigma}(T_{rim})} > 1$, has $\psi\sim 1$. Note that
for gray dust this makes equation \ref{Rcorrect} identical to the original DDN
expression (their equation 14). The approximate near-IR bump strength is given
in equation \ref{f2f1}, yielding
 \eq{
     {F_{2\mu m}\over F_{1\mu m}} \sim
   \cases{ 0.23       &   non-gray dust \cr
                                     \cr
           0.09+0.52\sin i &   gray dust \cr
    }
 }
for $T_*$=10,000 K, $T_{rim}$=1,500 K and $H_{rim}/R_{rim}$=0.2. Comparison of
this result with the data in figure \ref{F2F1_Dullemond} shows that the NIR
bump of non-gray dust is too small to explain the observations. Therefore,
interpretation of the NIR bump in Herbig Ae stars with inner disk puffing
places a strong constraint on dust evolution in this region. The dust must grow
to a size greatly exceeding the initial interstellar size distribution, and
small grains must be depleted to such a large extent that the inner disk
opacity can be considered gray. In the next subsection we employ exact 2D
radiative transfer code to obtain accurate values for $\psi$ and place more
precise constrains on the DDN model.

\subsection{Exact Models for Single-Size Grains}
\label{exact_DDN}

To examine the validity of conclusions based on our approximate solution we
performed full 2D radiative transfer calculations for an optically thick torus
centered on a 10,000 K star. The torus cross section is sketched in figure
\ref{DDN_sketch}; it is a square with side-length of $2H_{rim}$, where
$H_{rim}=0.2R_{rim}$. This configuration is the same as described by DDN, where
the puffed-up disk rim is a cylindrical surface directly exposed to stellar
radiation, while the rest of the inner disk is in its shadow. The dust has
sublimation temperature $T_{sub}=$1,500 K and constant density everywhere in
the torus, with horizontal and vertical optical depths $\tau_V=10,000$ in
visual. Different density structures do not change our results as long as the
$\tau_V=1$ layer on the illuminated surface is geometrically much smaller than
$H_{rim}$.

Radiative transfer modeling was conducted with our code LELUYA
(http://www.leluya.org) that works with axially symmetric dust configurations.
It solves the integral equation of the formal solution of radiative transfer
including dust scattering, absorption and thermal emission. The solution is
based on a long-characteristics approach to the direct method of solving the
matrix version of the integral equation \citep{Kurucz69}.

The results are shown in figure \ref{LELUYAmodels} together with the original
DDN solution (dashed line). Our 2D model results for gray dust without
scattering (solid line) are very close to the \citet{Dullemond2002} results,
shown in figure \ref{F2F1_Dullemond}, which also included vertical hydrostatics
equilibrium.  This model has $\psi=$4. Its rim radius (49$R_*$) and flux are
essentially the same as the original DDN model, confirming that a puffed-up rim
of gray dust is capable of explaining the near IR bump.

Since realistic dust is not gray at all wavelengths, we calculated models for
silicate dust with different grain radii, employing optical constants from
\citet{Dorschner95} ($x$ = 0.4 olivine). Figure \ref{LELUYAmodels} shows
results for three representative grain radii, with the corresponding rim
properties summarized in table \ref{table-grains}. Our model results for
0.1$\mu$m grains are almost identical to the \citet{Dullemond2002} results for
purely small grains (see figure \ref{F2F1_Dullemond}). As is evident from
figure \ref{LELUYAmodels}, the model can explain the data when the grain radii
are 2$\mu$m and 0.5$\mu$m, but it starts to fail as a general explanation of
the near IR bump when the grain radius drops below $\sim 0.1\mu$m.

\begin{figure}
 \plotone{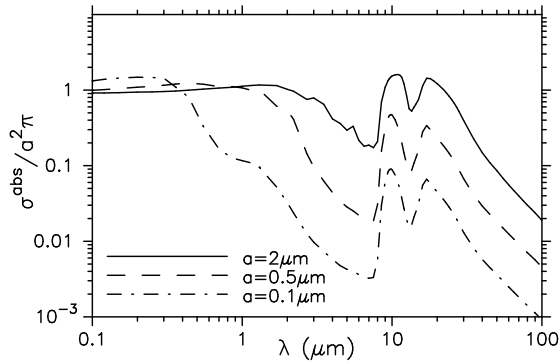}
 \caption{ \label{Cross_section}
Dust absorption cross sections for the three grain radii used in the exact
radiative transfer calculations shown in figure \ref{LELUYAmodels}.
}
\end{figure}

A decrease in grain size has two opposing effects on the rim flux. On one hand,
the ratio $\bar{\sigma}(T_*)/\bar{\sigma}(T_{sub})$ is increasing, leading to a
larger rim radius and emitting area and thus enhancing the rim emission. On the
other, the rim surface brightness is declining because $q_{\lambda}$ is
decreasing, reducing the rim emission. The net result is that maximum rim
emission occurs at grain radius of $\sim$0.5$\mu$m, which, as is evident from
figure \ref{Cross_section}, corresponds to the transition between gray and
non-gray opacity in the near IR region.  This is predominantly a grain
size effect; the dust chemistry introduces only second order corrections.

When the grain radius drops below 0.5$\mu$m the dust opacity becomes non-gray in
the near IR and the puffed-up rim model begins to fail. The flux of the
0.1$\mu$m grain model, which is almost angle independent as predicted by
equation \ref{f2f1}, can reproduce only the weakest near-IR bumps. Therefore
the DDN model can explain the near IR bump in Herbig Ae stars only when {\em
both} of the following conditions are met:

1) {\it the rim dust opacity is gray in the near IR (grain radius $\ga$
0.5$\mu$m)},

and

2) {\it the disk is puffed to a height $H_{rim}/R_{rim}\ga 0.15$}.

\noindent Figure \ref{F2F1_Dullemond} shows that for these conditions
to be satisfied, the DDN model requires the complete absence of small
grains in the disk inner region. Therefore, for this model to work,
the rim dust must undergo substantial growth that also fully depletes
the population of small grain. At the same time, this process cannot
be so extreme in the rest of the disk because the mid IR spectrum of
Herbig Ae stars displays the dust features of small grain emission
\citep{Boekel05}.

\begin{deluxetable}{cccccc}
\tabletypesize{\scriptsize}
 \tablecaption{\label{table-grains}
The exact single-grain DDN model results
}
\tablewidth{0pt}
\tablehead{
\colhead{Grain radius} &
\colhead{$R_{rim}/R_*$} &
\colhead{$\bar{\sigma}(T_*)/\bar{\sigma}(T_{sub})$} &
\colhead{$\psi$} &
\colhead{$q(2\mu m)$}
}
\startdata
gray      & 49  & 1.0 & 4.0 & 1.00 \\
2$\mu$m   & 52  & 1.3 & 3.6 & 0.99 \\
0.5$\mu$m & 68  & 3.6 & 2.3 & 0.45 \\
0.1$\mu$m & 150 & 28  & 1.3 & 0.10 \\
\enddata
\end{deluxetable}

\begin{figure*}
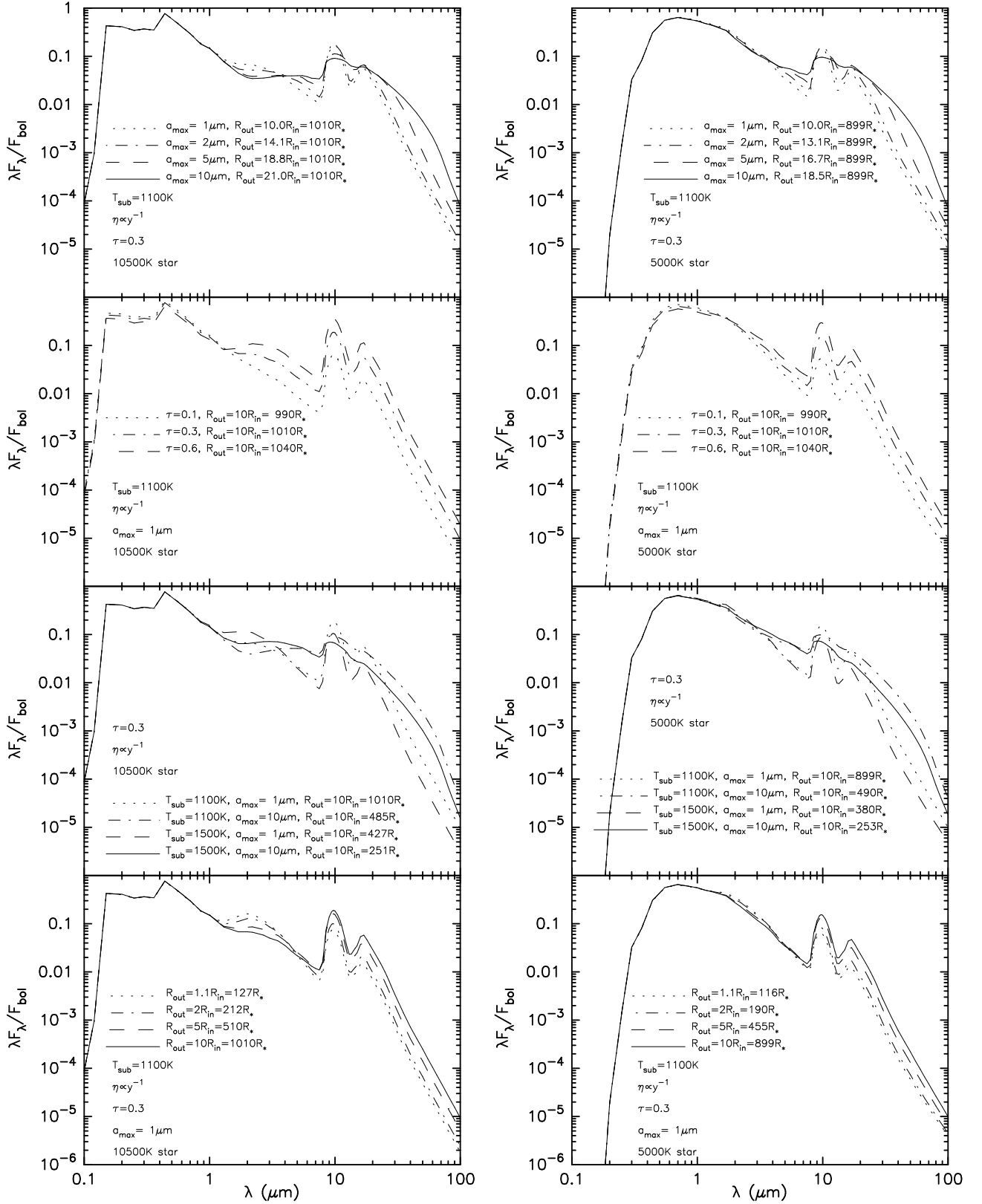

 \epsscale{.55}
 \plotone{f5a.eps}
  \epsscale{.55}
 \plotone{f5b.eps}
 \caption{\label{fig_SED1}
SED variation with the parameters of small spherical halos around stars with
10,500 K and 5,000 K (Kurucz stellar models). The radial density profile is
$\eta\propto y^{-1}$. The dust chemistry is $x$ = 0.4 olivine from
\citet{Dorschner95}. The grain size distribution is n($a$)$\propto a^{-2}$
between minimum grain radius 0.01$\mu$m and maximum $a_{max}$, as marked. The
other varied parameters are the dust sublimation temperature $T_{sub}$, which
sets the halo inner radius $R_{in}$, the halo outer radius $R_{out}$ and its
optical depth at visual $\tau$. Note the variation of $R_{in}$ among models,
especially when $a_{max}$ is increased.
}
\end{figure*}

\begin{figure*}
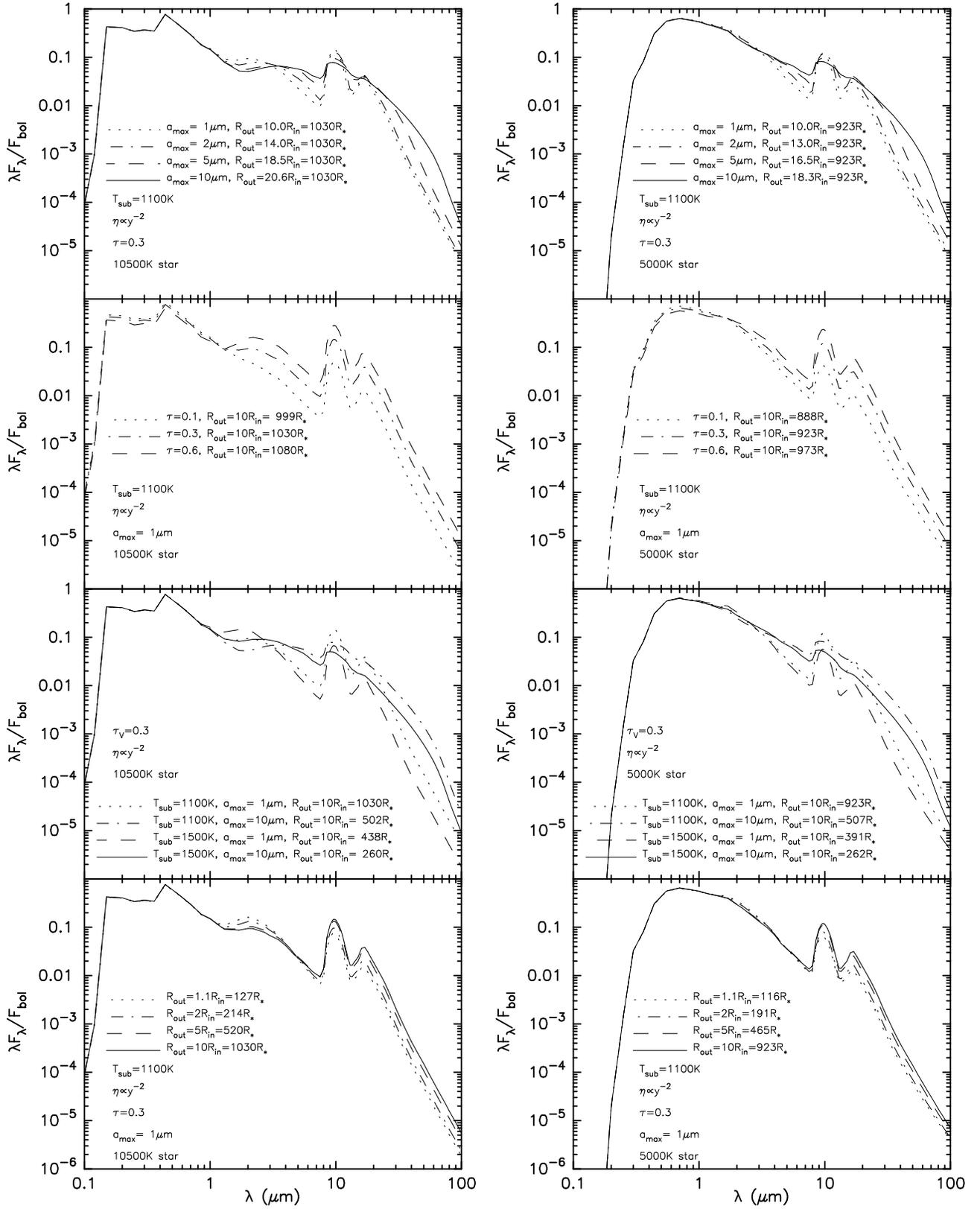

  \epsscale{.55}
\plotone{f6a.eps}
  \epsscale{.55}
\plotone{f6b.eps}
\caption{\label{fig_SED2}
The same as figure \ref{fig_SED1}, except that $\eta\propto y^{-2}$.}
\end{figure*}

\begin{figure}
 \plotone{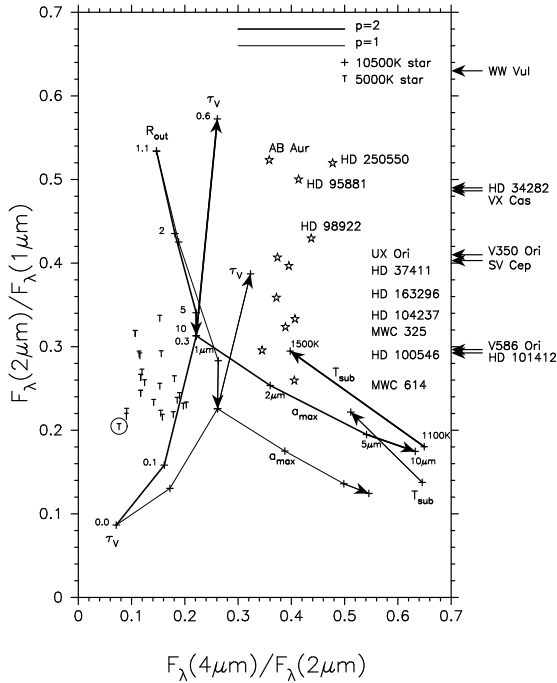}
\caption{\label{halo_bump} Diagram of the strength (vertical axis) vs.\ shape
(horizontal axis) of the near IR bump. Data are for the same sources as in
figure \ref{F2F1_Dullemond}. Objects with enough data to determine the bump shape
are marked with {\it stars}, otherwise only their bump strengths are marked
with arrows on the right. Theoretical results for the halo models presented in
figures \ref{fig_SED1} and \ref{fig_SED2} are marked with {\it crosses} for
halos around a 10,500K star and with {\it T\/} for halos around a 5,000K star
(the location of this naked star is marked with encircled T). Model results for
the 10,500K case are connected with arrowed lines that indicate their path in
the diagram when only one parameter is varied, as marked. Thick lines
correspond to $p$ = 2 halos, thin lines to $p$ = 1.
}.
\end{figure}

\section{The near IR bump and imaging explained with a dusty halo}

A dusty halo around the disk inner regions ($\sim$ 10 AU) has been
invoked to explain polarimetric measurements \citep{Yudin} and
correlations between variabilities in the optical and near IR
\citep{Eiroa}.  Such small regions are not yet accessible to direct
imaging but have been resolved in near IR interferometry by
\citet{Millan-Gabet-survey} who also favor the halo geometry, although
the interpretation of these visibility data is still model-dependent.
Direct imaging is currently available only for larger scales, and
these observations have revealed larger halos, $\ga$ 100 AU, around
some objects (V03). The relation between the inner and outer halos,
whether they are simply the inner and outer regions of the same
circumstellar component, remains an open question. However, at the
phenomenological level this issue is not relevant because the two can
be treated as separate circumstellar components if both are optically
thin. The inner halo is then radiatively decoupled from the cooler
outer halo, simplifying the study of inner halos.

Here we explore the contribution of the inner halo to the near IR emission. The
halo precise geometry is not particularly important. It could be elongated,
clumpy or inhomogeneous, but as long as it is optically thin it can be
approximated with spherical geometry. The reason is that the temperature of
optically thin dust is dominated by the stellar heating, resulting in
spherically symmetric isotherms and circularly symmetric images at wavelengths
where the dust thermal emission dominates over scattering (V03). Optically thin
halos are also transparent to the disk emission and we can ignore the disk
effect on the halo. The exact image shape ultimately depends on detailed
dust density and grain properties, telescope resolution and sensitivity,
observational wavelength and the intrinsic ratio between the disk and halo
surface brightness. Various observations of {\it R Mon} vividly illustrate
these effects \citep[see][]{Weigelt}.

If the halo optical depth at visual wavelengths $\tau_V$ is larger
than $\case14 H/R$, where $H/R$ is the disk flaring at the halo outer
radius, then the halo dominates the SED coming from the dust within
radius $R$ around the star (see V03 for details).  At near IR
wavelengths, this condition is satisfied for the halo optical depths
of interest here ($\tau_V\ga 0.1$).

\begin{figure}
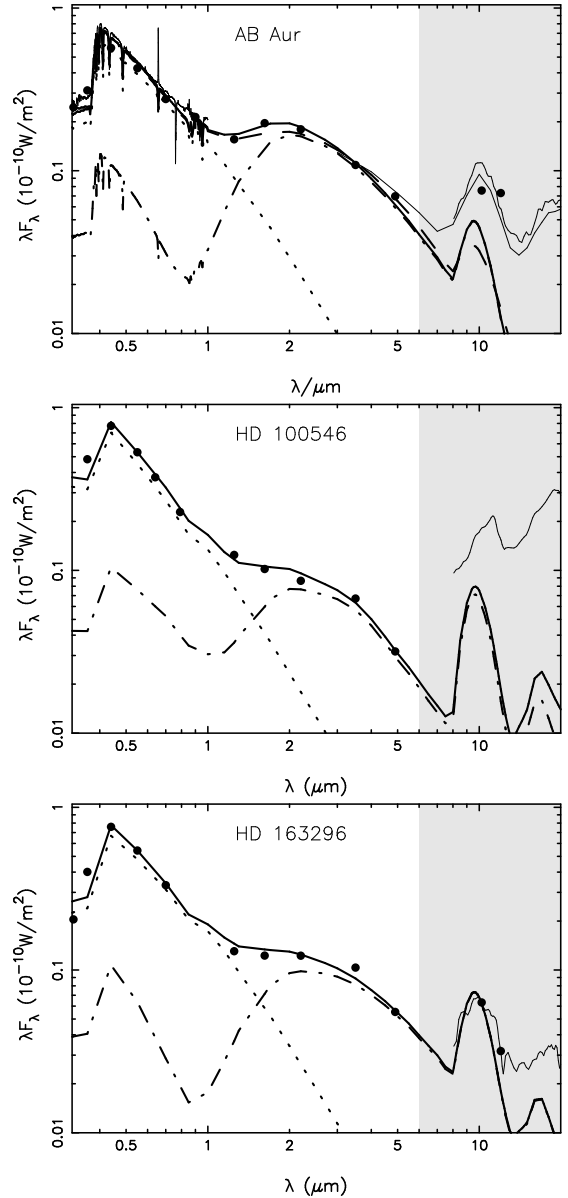

 \plotone{f8a.eps}\\
 \plotone{f8b.eps}\\
 \plotone{f8c.eps}
\caption{\label{fig_objects} Models for the near IR bumps in AB Aur, HD 100546
and HD 163296. Points and thin solid lines are the data (see
\S\ref{section2}). Measurement uncertainties in the near IR are comparable to
the symbol sizes. Thick solid lines are the models, comprised of the stellar
component (dotted lines) and a halo contribution (dash-dotted lines); see text
and table \ref{table-fit} for model details. In the case of AB Aur, the thick
dashed line is the second model listed in table \ref{table-fit}. Only data at
wavelengths shorter then 6$\mu$m were employed in the fits since in this study
we consider only the inner halo responsible for the near IR bump. Longer
wavelengths (the gray area) are displayed to show the halo contribution to the
mid IR features; in HD 163296, this contribution suffices to explain the
observed 10\mic\ feature.
}
\end{figure}

\begin{figure}
 \epsscale{1.15}
 \plotone{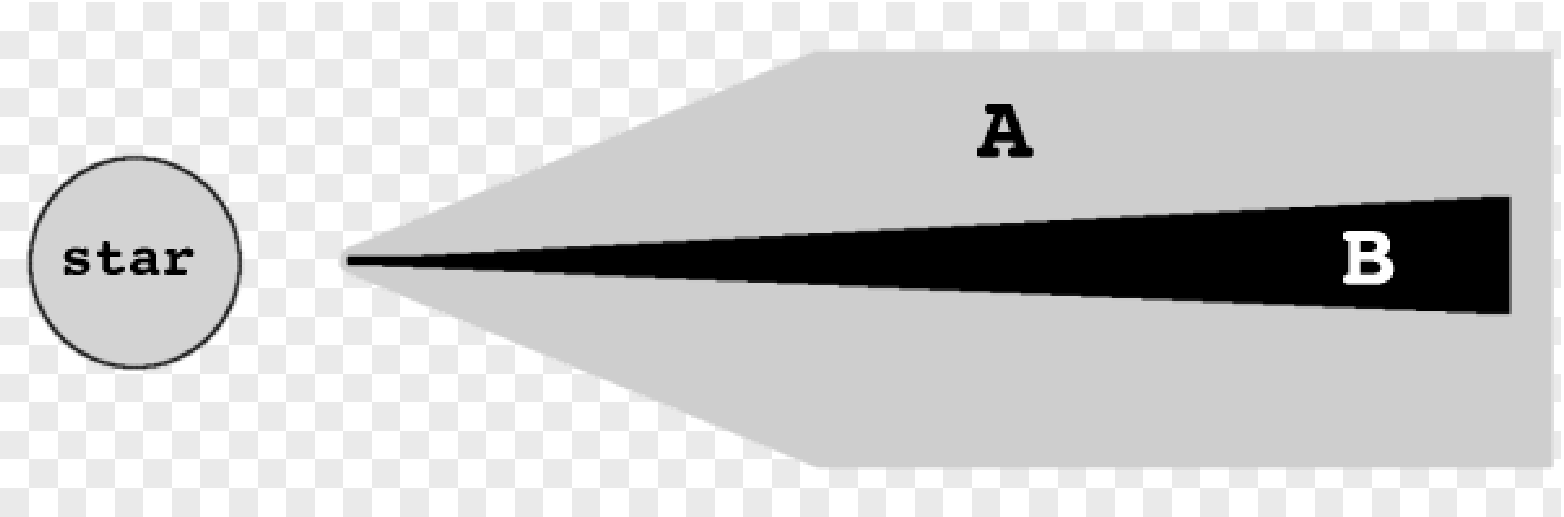}
\caption{\label{hd100546sketch} Sketch of the circumstellar geometry indicated
by imaging observations of HD 100546 (\citet{Grady01}, \citet{Grady05}; see
\S\ref{sec:HD 100546}). Region A is an optically thin dust layer, dominating
the near IR flux. The disk, marked with B, is cooler and does not affect the
near IR wavelengths. Both regions A and B are embedded in an optically thin
dusty envelope, marked with {\it checkered pattern}, whose optical depth is so
small that its contribution, too, to the near IR flux is negligible.
}
\end{figure}

\subsection{Theoretical Examples}
\label{various_models}

Our models consist of a star surrounded by a spherical halo with
radial density profile $\eta\propto r^{-p}$. The halo extends from
inner radius $R_{in}$, set by the dust sublimation temperature
$T_{sub}$, to outer radius $R_{out}$.

The dust chemistry is $x$=0.4 olivine from \citet{Dorschner95}, with
grain size distribution n($a$)$\propto a^{-q}$ between the minimum grain
radius $a_{min}$ and maximum $a_{max}$. We use $q=2$, $a_{min}=0.01\mu$m
and vary $a_{max}$. The radial optical depth of the halo is specified
at $\lambda$ = 0.55$\mu$m as $\tau_V$. The radiative transfer problem
is solved with the code DUSTY \citep{DUSTY}, which takes advantage of
the scaling properties of the radiative transfer problem for dust
absorption, emission and scattering \citep{IE97}.

Figures \ref{fig_SED1} and \ref{fig_SED2} show some SED examples for dusty
halos around 10,500K and 5,000K stars, representative of Herbig Ae and T
Tauri stars, respectively. The stellar spectrum is taken from Kurucz models.
In addition to the strength parameter $F_{2\mic}/F_{1\mic}$, the flux ratio
$F_{4\mic}/F_{2\mic}$ can be used to characterize the NIR bump shape. Both the
strength and shape parameters are influenced by changes in the dust sublimation
temperature, maximum grain size, halo outer radius and optical depth.
Comparison of the data with halo model results for the strength and shape
parameters is shown in figure \ref{halo_bump} for the same objects as in figure
\ref{LELUYAmodels}. Models for $p = 1$ and $p = 2$ halos around a 10,500K star
are dispersed all over the diagram. Arrows show how the model results move in
the diagram as the model parameters are varied, indicating that various
degeneracies are possible. The observed levels of bump strength and shape are
readily reproduced with plausible values of the model parameters.

We briefly summarize the effect of various halo parameters on the strength and
shape of the near IR bump.

{\it Optical depth}: A larger optical depth results in a stronger near IR bump.
This reflects the dependence of flux on the total mass of emitting dust
(equations A7 and A12 in V03). The dust sublimation radius
$R_{in}$ is only slightly affected, as expected in the optically thin limit where
the diffuse radiation is negligible.

{\it Grain size}: Larger grains shift the near IR bump toward longer
wavelengths and make it appear more flat. With increased grain size the opacity
becomes more similar to gray dust, resulting in a $r^{-0.5}$ temperature
profile since the geometrical dilution of stellar heating is the only cause of
temperature variation. Smaller grain sizes create steeper radial temperature
profiles. Therefore, for a given density profile smaller grains emit relatively
more radiation at shorter wavelengths than larger grains. In practice, grain
sizes come in mixtures and sublimate at different radial distances, greatly
adding to the complexity of the problem. The SED models are therefore prone to
various model degeneracies.

{\it Sublimation temperature}: With a higher dust sublimation temperature, the
near IR bump shifts to shorter wavelengths, reflecting the shift of the
emission peak.

{\it Outer radius}: The halo size can affect the near IR bump in two ways. On
one hand, reducing the outer radius while keeping the dust distribution fixed
reduces also the total optical depth. The near IR bump then starts to decrease
when the dust removal reaches the NIR emission regions at radial distance $\sim
10 R_{in}$ (temperatures $\ga$ 500 K). On the other hand, reducing the outer
radius at a fixed halo optical depth is equivalent to redistributing the dust
within the halo. The bump then becomes stronger as the outer radius is reduced
because more dust is shifted toward smaller radii and higher temperatures.

{\it Stellar temperature}: As its temperature decreases, the emission from the
star starts to blend with that from the halo, and the near IR bump disappears.
Only a careful analysis can then separate the stellar from the diffuse flux in
the near IR and reveal the bump. For comparison with T Tau stars, figure
\ref{halo_bump} shows also 5,000 K models (marked with {\it T}). In spite of
the large variations in halo parameters, these models display only a limited
range of bump strengths and shapes close to the naked star values. This
explains why the near IR bump was not originally recognized in T Tau stars
while easily detected in Herbig Ae/Be stars.

\begin{deluxetable*}{ccccccccccc}
\tabletypesize{\scriptsize}
\tablecaption{\label{table-fit} Halo model parameters for three case studies of
Herbig Ae stars
}
\tablewidth{0pt}
\tablehead{
\colhead{Object} &
\colhead{T$_*$} &
\colhead{A$_V$} &
\colhead{T$_{sub}$} &
\colhead{$\tau_V$} &
\colhead{$\eta$} &
\colhead{$R_{in}$} &
\colhead{$F_{bol}$} &
\colhead{grain radius} &
\colhead{carbon} &
\colhead{olivine}
}

\startdata
AB Aur    &  9750 & 0.35 & 1500 & 0.35 & r$^{-2}$ & 0.78 & 0.9 & $q=3.5$,  $a_{max}=0.25$ & 40\% & 60\% \\
AB Aur    &  9750 & 0.35 & 1800 & 0.35 & r$^{-2}$ & 0.25  & 0.9 & $q=3.5$,  $a_{max}=5.0$ & 40\% & 60\% \\
HD 100546 & 10500 & 0.3  & 1500 & 0.35\tablenotemark{a}
                                & r$^{-1}$ & 0.45 & 0.8 & $q=2$,   $a_{max}=0.50$ & 10\% & 90\% \\
HD 163296 &  9500 & 0.3  & 1500 & 0.25 & r$^{-1}$ & 0.61 & 0.9 & $q=3.5$,  $a_{max}=0.25$ & 30\% & 70\% \\
\enddata

 \tablecomments{
Description of columns: T$_*$ --- stellar temperature in K. T$_{sub}$ --- dust
sublimation temperature in K. A$_V$ --- reddening toward the star by dust other
than the inner halo. $\tau_V$ --- visual optical depth of the halo. $\eta$ ---
radial dust density profile of the halo. $R_{in}$ --- halo inner radius in AU,
determined from dust sublimation; the halo outer radius is 10$R_{in}$. $F_{bol}$
--- total bolometric flux in 10$^{-10}$W/m$^2$. grain radius --- dust size
distribution $a^{-q}$ between minimum $a_{min} = 0.005$ $\mu$m and maximum
$a_{max}$ listed in $\mu$m. Amorphous carbon properties from \citet{amC},
olivine from \citet{Dorschner95} (with x = 0.4).
}
\tablenotetext{a}{HD 100546 is modeled with a flattened halo, which does not
contribute to the circumstellar reddening because the dust is out of the line
of sight toward the star (see figure \ref{hd100546sketch}).}
\end{deluxetable*}

\subsection{Observational Examples}
\label{sec-examples}

We show three examples which illustrate different circumstellar dust
configurations: AB Aur, HD 100546 and HD 163296. HST imaging suggests that AB
Aur and HD 100546 have large halos at radii $\ga 1$'', while HD 163296 shows
only a disk \citep{Grady03}. Irrespective of the existence of a large halo, all
three objects show a near IR bump, with the strongest bump in AB Aur. Since the
focus of this study is the near IR bump, the large scale halos are irrelevant
here and we only consider a small halo within $\sim$10 AU around the star.

Our fits to the data are shown in figure \ref{fig_objects}, with the model
parameters listed in table \ref{table-fit}. The halo outer radius is 10 times
the dust sublimation radius in all models. Since our focus is the near IR bump,
our model consists only of the star and the inner halo, and only data at
wavelengths shorter then 6$\mu$m were employed in the fitting. The derived
model parameters are not unique since various degeneracies exist in model
results for the near IR flux (see \S\ref{various_models}). For example, the
``hot component" in the \citet{Bouwman} models can be interpreted as a
small-scale halo with dust properties different from those in our study.

\subsubsection{AB Aur}

The emission from AB Aur has been resolved at various wavelengths and
interpreted as a disk with vastly different estimates for the inclination
angle, as follows:

\begin{tabular}{l l l}
                    \\
 {\em visual\/}:    & $i\la 45\deg$        &\cite{Grady99}             \\ \\
 {\em near IR\/}:   & $i\la 30\deg$        &\cite{Eisner03}, (2004)    \\
                    & $i=30\pm 5\deg$      &\cite{Fukagawa_AB_Aur}     \\ \\
 {\em mid IR\/}:    & $i=55\pm 10\deg$     &\cite{Liu04}               \\ \\
 {\em millimeter\/}:& $i=17\deg^{+6}_{-3}$ &\cite{Semenov04}           \\
                    & $i=21\deg.5^{+0.4}_{-0.3}$ &\cite{Corder}        \\
                    & $i=33\pm 10\deg$     &\cite{Pietu}               \\
                    & $i\sim 76\deg$       &\cite{MS97}                \\
                    \\
\end{tabular}

Such a disparity is expected in halo-embedded disks (see figure 7
in V03) because, as noted by \citet{MIVE}, the halo dominates the
images at wavelengths extending to $\sim$ 100 \mic\ or so, and the disk emerges
only at longer wavelengths. Interpretation of molecular line images, too, must
be done carefully to avoid confusion between the halo and disk contributions.

A general conclusion about the AB Aur inner halo is that it must have a radial
density profile between $1/r$ and $1/r^2$;  this differs from the outer
halo, which has a $1/r$ density profile as deduced from the $1/r^2$ radial
brightness profile of the HST image \citep[][ see also equation A10 in
V03]{Grady99}. Conclusions regarding the properties of the dust grains in
the inner halo are less firm. Near IR interferometry suggests the presence of
dust close to the star, implying large grains that can survive at small
distances. An example of a big grains model for AB Aur is shown in figure
\ref{fig_objects} with thick dashed line (see also table \ref{table-fit}). The
grain size and chemistry might be subject to radial variations, as is indicated
by comparison between the HST \citep{Grady99} and Subaru images
\citep{Fukagawa_AB_Aur}, further complicating the modeling.

\subsubsection{HD 100546}
\label{sec:HD 100546}

The HST image of this source \citep{Grady01} shows a very tenuous large scale
nebulosity, whose low surface brightness implies an optical depth of only
$\tau_V \sim$ 0.015. This component of the dust distribution does not
contribute significantly to the IR emission and can be ignored in the current
analysis. The HST image, which is produced purely by scattered light, reveals
also a prominent disk with inclination angle $49\pm4\deg$. Near IR
\citep{Augereau} and mid IR \citep{Liu03} imaging give similar results for the
disk even though the latter is produced purely by dust emission and the former
contains a mixture of both scattering and emission. The HST brightness contours
are symmetric, with the brightness declining as $1/r^3$. These are the
signatures of scattering from the CG layer of a flat disk (see V03). However,
for the CG model to be applicable, every point on the scattering surface, which
extends to a distance of $\sim$10\arcsec\ from the star, must have a direct
line of sight to the stellar surface. This is impossible in the case of a flat
disk, since it would have to maintain a thickness smaller than the stellar
radius for hundreds of AU. Therefore, the only self-consistent explanation of
the HST imaging is with an optically thin halo whose dimensions are unrelated
to the stellar size. The HST image implies that the halo has a flattened
geometrical shape, and its $1/r^3$ brightness profile implies that it has a
$1/r^2$ radial density profile (V03). This flattened halo is outlined as region
A in figure \ref{hd100546sketch}. Since the halo dominates the imaging, the
geometry of the optically thick disk structure, outlined as region B in the
figure, remains unknown.

The HST imaging does not constrain the inner-halo geometry at radii $\la$ 10
AU.  The surface density must be reduced in that region because the near IR
bump in HD 100546 is significantly smaller then in AB Aur even though otherwise
the two stars are rather similar. Indeed, the fit to the near IR bump yields a
$1/r$ radial density profile (figure \ref{fig_objects} and table
\ref{table-fit}), shallower than in the region resolved by HST. The fit was
further improved by an increased contribution from large grains ($a_{max} = 0.5
\mu$m and $q = 2$) and a reduced fraction of carbon dust in the mix.
Observations by \citet{Grady05} show that a constant density profile, creating
$1/r$ brightness profile, might be more appropriate in the region between 20
and 50 AU. We find that a constant density model could also fit the near IR
spectrum if the sublimation temperature were increased to 1700 K. All these
results point toward large structural differences between the inner and outer
regions of HD 100546.

\subsubsection{HD 163296}

The model properties of the inner halo in this source are very similar
to AB Aur, except that a shallower density profile of $p=1$ is
preferred (see table \ref{table-fit}). A similar general conclusion is
that the halo radial density profile is between $1/r$ and $1/r^2$,
with uncertainties in the dust properties. Significantly, in this
source the inner halo also fits the 10$\mu$m feature all by itself
(figure \ref{fig_objects}). No other optically thin components are
required for explaining the mid IR dust features, indeed none are
observed. The HST image, which is incapable of resolving the inner
halo, shows no evidence of a large scale structure other then the
disk, with an inclination of $60\pm5\deg$ \citep{Grady00} in agreement
with $\sim 58\deg$ derived from millimeter observation
\citep{MS97}. Another noteworthy feature of the HST image is a bipolar
jet. The process responsible for jet formation could perhaps also lift
up dust above the disk and create the small scale halo responsible for
the near IR bump. Such possible correlation between jets and the near
IR bump can be studied further when more high resolution data from a
larger sample of objects become available.

\begin{figure}
 \epsscale{1.15}
 \plotone{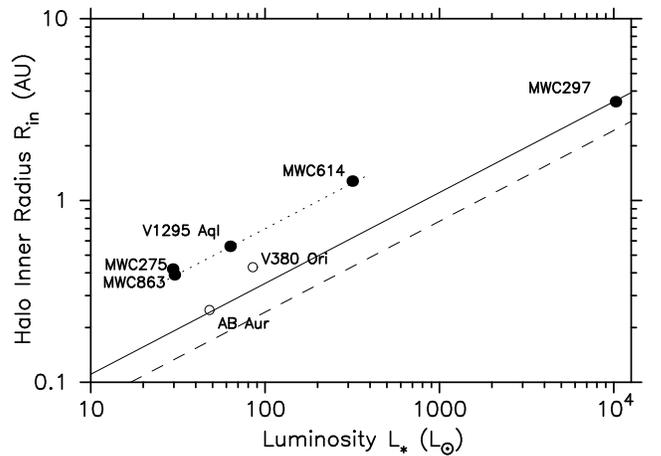}
 \caption{\label{Sublimation_radius}
Size-luminosity diagram for the inner-halo model. The sublimation radius $R_{in}$
of each displayed object is determined from a simultaneous fit to its near-IR
bump and visibility data. {\it Filled circles} mark objects with dust
sublimation temperature 1500 K, {\it empty circles} with 1800 K. Lower limits
on $R_{in}$ are shown with {\it solid line} for $T_{sub}$ = 1500 K and {\it dashed
line} for $T_{sub}$ = 1800 K; they correspond to halos of gray dust and zero
optical depth. The {\it dotted line} connects a group of objects that follow
closely the $R_{in}\propto L_*^{1/2}$ relation, indicating that their inner halos
have similar optical depth and dust properties. The original version of this
diagram, constructed from ring-model visibility analysis, produced a puzzling
amount of scatter \citep[ see \S\ref{sec:R1-L}]{Monnier}. The inner-halo model
resolves the puzzle.
}
\end{figure}

\subsection{Size-luminosity Correlation}
\label{sec:R1-L}

The milli-arcsecond resolution reached in near IR interferometry enables
studies of the immediate environment of young stars, down to 0.1 AU
\citep{Malbet}. Unfortunately, current visibility data are not yet capable of
reproducing the full 2D image of an object, instead requiring a model of the
geometry for their analysis. One simple and often used model of the
circumstellar geometry is a flat dust ring of uniform surface brightness. This
ad hoc model did not arise from some specific radiative transfer modeling but
rather chosen as a simple approach to the visibility fitting procedure.

Fitting the visibility data of a number of objects with this ring model,
\citet{Monnier} discovered that the size of the ring inner radius increased
with the stellar luminosity $L_*$. This is the expected result when dust
sublimation controls the size of the dust-free region around the star. Since
radiative transfer is scale invariant \citep{IE97}, inner radii of rings would
be expected to scale as $L_*^{1/2}$ if their dust properties were the same.
However, \citet{Monnier} do not find such a trend. Instead, at a fixed
luminosity the derived radii vary by almost a factor of ten, which they refer
to as scatter in the size-luminosity diagram. This scatter indicates either
that the disk inner regions have vastly different properties, with the
sublimation temperature varying from $\sim$1000 K to $\sim$2000 K, or that the
ring model is not a proper description of the actual dust distribution.
\citet{Monnier} also noted that some highly luminous objects ($L_*\ga 10^3
L_{\odot}$) had smaller than expected inner ring radii, thus requiring even
higher dust sublimation temperatures. New interferometric data by
\citet{Monnier05} slightly reduce the scatter in the ring-radius--luminosity
relation, but the remaining scatter still implies a large range of sublimation
temperatures, and very luminous objects still display abnormally small radii.

Instead of the ring model we have analyzed the interferometry results
with the inner-halo model, performing simultaneous fits of both the
near-IR bump and visibility data. Preliminary results are shown in
figure \ref{Sublimation_radius}.  It is highly significant that there
are no objects in the forbidden region below the indicated lower
limits. The correlation of overall bolometric luminosity with inner
radius is much tighter than in the ring model, the small remaining
scatter arises from variation in halo optical depth and grain size.
In contrast with the ring model, the sublimation temperature rarely
differs from 1500K (it is 1800K in a couple of objects).  The high
luminosity object MWC297 which was especially troubling in the
\citet{Monnier} analysis is now consistent with 1500K sublimation
temperature.  It is striking how some of the objects that were highly
scattered in the diagram by \citet{Monnier} now settle on the same
$L_*^{1/2}$ size-luminosity relation (dotted line in figure
\ref{Sublimation_radius}), indicating similarities in the halo
properties of all these stars, in turn pointing toward a common
physical mechanism of halo formation.

\section{Conclusions}

An examination of the puffed-up disk rim model (DDN) shows that it
has rather limited capabilities in explaining the near IR bump of
Herbig Ae/Be stars. The observed level of near IR excess implies a
certain emitting volume of optically thin puffed-up disk rim
surface for given dust properties. The volume derived from the DDN
model falls short of this observational limit, unless the disk is
made of perfectly gray dust. The puffed-up rim produces enough
near IR flux {\em only} when the inner disk consists purely of
dust grains larger than $\sim$ 0.5 $\mu$m {\it and} the disk
puffing reaches values of $H_{rim}/R_{rim}\ga 0.15$.  Models by
\citet{DD04} show that even traces of small grains inhibit the
disk puffing, eliminating the DDN model as a viable explanation of
the near IR bump.  Since the 10$\mu$m emission feature indicates
the presence of small grains in the circumstellar dust, additional
mechanisms must be invoked to remove all small grains from the
inner disk and keep the DDN model viable.

From fits to the SED of a number of Haebes, \citet{Dominik} conclude that the
infrared excess in these stars is produced by disks alone without the need for
additional circumstellar components. This conclusion is invalidated by the
mathematical proof that a fit to the SED cannot distinguish between the surface
of a flared disk and an optically thin halo (V03). Fits to the SED alone are
not a conclusive proof of a particular dust geometry.

We find that that the optically thin dusty halos around the disk
inner regions whose existence has been inferred in various
observations readily explain the strength and shape of the near IR
bump. The halo is not limited by the disk properties. Hence, it
can extend above the disk surface and accommodate the emitting
optically thin dust volume required by the near IR flux
observations. The required halo is rather small, less than several
AU in size, and its optical depth in visual is less than $\sim$
0.4. Despite its small optical depth, the halo dominates the near
IR spectrum and hides the disk near IR signature. However,
detailed properties of the halo, such as its exact shape, grain
properties or dust density profile, are not uniquely constrained
by the SED since different combinations of the parameters can
produce the same flux. These degeneracies can be broken only with
imaging capable of resolving the disk inner regions.

Inner halos not only explain the near IR bump but also successfully resolve the
puzzle presented by the relations between luminosities and near-IR
interferometric sizes \citep{Monnier}. In addition to their near IR emission,
the halos contribute also to the mid IR flux. HD 163296 is an extreme example
where the halo in itself fully explains the mid IR dust features without the
need for additional extended components (figure \ref{fig_objects}). The absence
of such components in the HST image of this source is another success of the
inner-halo model. In general, though, the inner halo emission is not expected
to dominate the mid IR but still make a significant contribution that must be
included in fits to the overall SED for reliable modelling of the rest of the
circumstellar material. Recently \citet{Boekel} suggested that differences in
the strength and shape of the mid IR silicate feature in Haebes are evidence
for dust settling in the disk. However, these differences could instead reflect
halo evolution, with the most active stars showing the strongest mid IR
signature of the inner disk halo. High resolution imaging is necessary for
definite conclusions about evolution either of the dust or the circumstellar
disk. Such imaging will soon become available at the VLTI, which offers
milli-arcsecond resolution at near IR.

\acknowledgments

We thank C.~P Dullemond, C. Dominik and A. Natta for fruitful discussions on
the physics of the DDN model. We also thank A. S. Miroshnichenko for help
with the data compilation.  DV thanks B. Draine and R. Rafikov for useful
comments. Support by the NSF grant PHY-0070928 (DV) is gratefully acknowledged.
This work was also supported by National Computational Science Alliance under
AST040006 and utilized the NCSA's Xeon Linux Cluster. DV also thanks the
Institute for Advanced Study (IAS) for time on their Linux cluster. TJ
acknowledges the hospitality and financial support of IAS during his visit to
the Institute. ME acknowledges partial support by NSF and NASA.

\appendix

\section{Approximate solution for the rim radius}
\label{appendixA}

The observed emission from a puffed up inner disk depends on the rim radius,
height and surface brightness (see \S\ref{approximate}). Deriving the rim
radius requires a proper treatment of the rim temperature structure. As already
noted by \citet{Dullemond2002}, in gray dust the diffuse radiation creates a
temperature inversion --- the dust temperature is maximum in the rim interior
(that is, at $R>R_{rim}$), not on the rim surface. Here we derive an
approximate solution for the dust temperature $T_0$ on the rim surface and the
temperature $T_1$ inside the rim at depth $\tau_V\sim1$ from the surface. The
solution demonstrates the inversion effect for gray dust and shows that it does
not exist in the non-gray case.

\subsection{Dust temperature $T_0$ on the rim surface}

Consider a dust grain on the rim surface. It is heated by the stellar flux
$F_*$ and the diffuse flux $F_{out}$ coming out from the rim interior. For
large optical depths these two fluxes are balanced: $F_*=F_{out}$. Stellar flux
absorbed by the grain is $\sigma_V F_*$, where $\sigma_V$ is the dust cross
section in visual. Absorbed diffuse flux is $\sim 2\sigma_{IR}F_{out}$, where
$\sigma_{IR}$ is the cross section in the near IR and the factor 2 accounts for
absorption from 2$\pi$ steradian. The grain emits into 4$\pi$ steradian, so
that the energy balance is
\eq{
  \sigma_V F_* + 2\sigma_{IR}F_{out} = 4\sigma_{IR}\sigma_{SB}T_0^4
}
where $\sigma_{SB}$ is the Stefan-Boltzmann constant. Using $F_*=F_{out}$ we
get
\eq{\label{T0}
  \sigma_{SB}T_0^4 = {F_* \over 4}\left(2+{\sigma_V\over \sigma_{IR}}\right)
}

\subsection{Dust temperature $T_1$ at $\tau_V\sim1$ from the surface}

Now consider a dust grain at distance $\tau_V\sim 1$ from the surface into the
rim. This grain is heated by the attenuated stellar flux
$F_*\exp(-\tau_V)=F_*\exp(-1)$, by the diffuse flux from the surface dust
between $\tau_V=0$ and $\tau_V\sim 1$, and by the diffuse flux from the rim
interior. The absorbed stellar flux is $\sigma_V F_*\exp(-1)$. Diffuse
contribution from the surface dust layer is $\sim
2\sigma_{IR}\sigma_{SB}T_0^4\tau_{IR}$, where $\tau_{IR}\sim
\sigma_{IR}/\sigma_V$ is the infrared optical depth of this surface layer.
Diffuse heating from the rim interior, described by temperature  $T_1$, is
$\sim 2\sigma_{IR}\sigma_{SB}T_1^4$. This is a good approximation for gray dust
and an overestimate for non-gray dust, where that temperature decreases rapidly
with optical depth. The energy balance is
\eq{
  \sigma_V F_*\exp(-1)+2\sigma_{IR}\sigma_{SB}T_0^4\tau_{IR}+
  2\sigma_{IR}\sigma_{SB}T_1^4 =  4 \sigma_{IR}\sigma_{SB}T_1^4
}
Using $T_0$ from equation \ref{T0} we get the interior temperature
\eq{\label{eq:T1}
  \sigma_{SB}T_1^4 = {F_*\over 4}\left(2{\sigma_{IR}\over \sigma_V}+1
                      +2{\sigma_V\over \sigma_{IR}}e^{-1}\right)
}
Note that the ratio $T_0/T_1$ depends only on $\sigma_{IR}/\sigma_V$ and is
independent of $F_*$.

\subsection{Gray and non-gray regimes}

We consider two distinct opacity regimes: gray when $\sigma_{IR}/\sigma_V\sim
1$ and non-gray when $\sigma_{IR}/\sigma_V\ll 1$. The ratio of the rim surface
temperature $T_0$ and the interior temperature $T_1$ in these two regimes is
\eq{
    T_0/T_1 = 0.95 \quad \hbox{when}\ \sigma_{IR}/\sigma_V=1
    \quad \hbox{(gray dust)}
}
\eq{
    T_0/T_1 \sim 1.08 \quad \hbox{when}\ \sigma_{IR}/\sigma_V\ll 1
      \quad \hbox{(non-gray dust)}
}
The gray opacity creates a temperature inversion with the temperature in the
rim interior higher than on its surface. This inversion does not appear in
non-gray dust where the temperature decreases monotonically with distance from
the rim surface. If the maximum dust temperature is 1,500K (sublimation
temperature) then gray dust has $T_1$=1,500 K and $T_0\sim$1,400 K, while
non-gray dust has  $T_0$=1,500 K and $T_1\la$1,400 K.

These approximate expressions are in reasonable agreement with the results
of exact 2D radiative transfer calculations (see \S\ref{exact_DDN}), which
yield
\[ \hbox{$T_0$=1387 K and $T_1$=1490 K} \quad \hbox{for gray dust} \]
\[ \hbox{$T_0$=1500 K and $T_1$=1236 K} \quad \hbox{for 0.1$\mu$m grains}\]
The transition between these two regimes occurs at grain radius 0.5$\mu$m which
yields  $T_0$=1474 K, $T_1$=1492 K and the maximum temperature of 1,500 K at
$\tau_V$=0.34.

\subsection{Disk rim radius and near-IR bump strength}

Since the rim optical depth is large, we can assume that both temperatures
$T_0$ and $T_1$ are located at essentially the same distance from the star. If
we set $T_1$ to the dust sublimation temperature $T_{sub}$ then based on
equation \ref{eq:T1} the rim radius is
\eq{
  R_{rim}={1\over 2}R_*\left({T_*\over T_{sub}}\right)^2
         \sqrt{2{\sigma_{IR}\over \sigma_V}+1
                      +2{\sigma_V\over \sigma_{IR}}e^{-1}}\sqrt{1+H_{rim}/R_{rim}}
}
where we used $F_*=L_*\sqrt{1+H_{rim}/R_{rim}}/4\pi R_{rim}^2$ (the factor
$\sqrt{1+H_{rim}/R_{rim}}$ is correction for rim self-irradiation, already
introduced by DDN). Comparison with equation \ref{Rcorrect} for the rim radius
gives
\eq{\label{psi}
  \psi = {2\over e}+ {\sigma_{IR}\over  \sigma_V}
          \left(1+2{\sigma_{IR}\over  \sigma_V}\right).
}
The two extreme opacity regimes yield
\eq{
\psi\sim
 \cases{ 1 & when $\sigma_{IR}/\sigma_V\ll 1$ (non-gray dust) \cr
                          \cr
         4  & when $\sigma_{IR}/\sigma_V \to 1$  (gray dust) \cr
        }
}
This result can also be derived by setting $T_0\sim T_{sub}$. Combining this
result with equations \ref{Fnongray} and \ref{Rcorrect} and dividing the
overall observed flux at 2$\mu$m by the stellar flux at 1$\mu$m yields the
near-IR bump strength
 \eq{\label{f2f1}
 {F_{2\mu m}\over F_{1\mu m}} =
   {B_{2\mu m}(T_*)\over B_{1\mu m}(T_*)} +
   \left({T_*\over T_{rim}}\right)^4 {B_{2\mu m}(T_{rim})\over B_{1\mu m}(T_*)}
   {H_{rim}\over \pi R_{rim}}
   \left(1+{H_{rim}\over R_{rim}}\right)\times
   \cases{ 1       &   non-gray dust \cr
                                     \cr
           4\sin i &   gray dust \cr
    }
 }
where we used the approximation $q_{\lambda}\sim  \sigma_{IR}/ \sigma_V \sim
\bar{\sigma}(T_{rim})/ \bar{\sigma}(T_*)$. This solution shows that non-gray
dust gives angle-independent bump strength in addition to reducing its
magnitude from the gray dust result.


\end{document}